# A Polynomial Time Algorithm for Graph Isomorphism

Reiner Czerwinski

Institut für Softwaretechnik und Theoretische Informatik
TU Berlin

August 20, 2018

**Abstract**

Algorithms testing two graphs for isomorphism known as yet in computer science have exponential worst case complexity. In this paper we propose an algorithm that has polynomial complexity and constructively supplies the evidence that the graph isomorphism lies in P.

## 1 Introduction

Graph isomorphism is a crucial problem in computer science and has been investigated intensely in the past. Applications can be found both inside and outside the computer science area. Symmetry breaking is an important example, which is a current topic in constraint solving and may be used for instance in register allocation, which is a critical performance issue. In chemistry, graph isomorphism algorithms are employed for molecule analysis.

Known algorithms like the ones presented by Nauty [1] and VF [2] do indeed indicate an instant isomorphism efficiently in the majority of cases. Yet, for each of these techniques, graphs leading to an exponential growth in calculation time can be constructed. For a long time, the existence of a method with polynomial runtime was in doubt, albeit Köbler [3],[4] had expressed that there were reasons to believe graph isomorphism not to be NP hard, unlike the subgraph isomorphism problem or the travelling salesman problem for which it is assumed that fast algorithms do not exist. He pointed out that the problem of finding an isomorphism is equally hard to the one of finding all isomorphisms which is unusual in NP-hard problems.

It has been well known [5] that a necessary (but not sufficient) criterion for graph isomorphism are identical eigenvalues of the adjacency matrix. This has been exploited [6] to develop an algorithm for the class of graphs with bounded eigenvalues.

The matrices will still have same eigenvalues, if you perturbs the diagonal elements of the matrices, such that a vertex in the first and in the second graph have same values for the diagonal elements, if they are candidates for an isomorphism. By iterative perturbation of the diagonal elements the two graphs can be checked for isomorphism in polynomial time.

A program for graph isomorphism with polynomial run time was already implemented by Trofimov and Smolenskii two years ago [7] by using this trick.

## 2 Terms and linear algebra

A graph $G = (V, E)$ consists of a set of vertices $V$ and a set of edges $E$. Two vertices $v1$ and $v2$ are connected or adjacent iff $\{v1, v2\} \in E$.

Two graphs $G = (V, E)$ and $G' = (V', E')$ are isomorph, iff a bijective mapping $\pi : V \to V'$ exists with $\{v1, v2\} \in E \Leftrightarrow \{\pi(v1), \pi(v2)\} \in E'$.

In the adjacency matrix an entry is 1 in the case when the two respective vertices are connected through an edge in $E$, otherwise 0. As we will only consider undirectional graphs, the adjacency matrix is symmetric. Köbler [4] has proven this to be sufficient as the graph isomorphism problem for directed graphs can be reduced to the one for undirected ones.

A symmetric matrix can be decomposed by means of eigenvalue decomposition into $A = V * D * V^T$, $D$ being a diagonal matrix, whose main diagonal entries are filled by the eigenvalues of the matrix A. $V$ is a unitary matrix, that is $V^T * V = I$. The columns of V are the eigenvectors of A. If $V = (v_1, ..., v_n)$ and $D = diag(\mu_1, ..., \mu_n)$, then $Au_i = \mu_i u_i$. $v_1, ..., v_n$ are orthonormal[8], p. 291, thus

$$<v_i, v_j> = v_i^T v_j = \begin{cases} 1 & \text{if } i = j \\ 0 & \text{otherwise} \end{cases}$$

.

Let $\{\mu_1, ..., \mu_m\}$ be the set of distinct eigenvalues. Associated with the eigenvalues $\mu_i$ is the eigenspace $S_i$ containing the eigenvectors associated with $\mu_i$ : $S_i = \{x \in R^n \mid Ax = \mu_i x\}$ . By virtue of the symmetry of A, we have:

1. If $\mu_i$ is an eigenvalue with multiplicity $m_i$ then $S_i$ has dimension $m_i$.

2. The direct sum $S_1 \oplus S_2 \oplus ... \oplus S_m$ equals $R^n$

3. If $i \neq j$ then $S_i$ and $S_j$ are mutually orthogonal.

## 3 Derivation

If two graphs G and G' are isomorphic, so for the respective adjacency matrices $A' = P * A * P^T$ must hold, where $P$ denotes a permutation matrix. This entails $A'^k = P * A^k * P^T$ for arbitrary k. $A'$ and $A$ have the same eigenvalues, because $P$ is an unitary matrix.

**Proposition 1.** *If A is an adjacency matrix of a graph and P is a permutation matrix, then P describes an automorphism of the graph, iff $PA = AP$.*

The proof is written in [5].

**Lemma 1.** *Let $A$ and $A' = P * A * P^T$ be adjacency matrices of isomorphic graphs and $D$ a diagonal matrix, so their characteristic polynomials are equal and $\chi(A+D) = \chi(A' + P*D*P^T)$ holds, i. e. the two matrices have the same eigenvalues.*

*Proof.* $A' + P * D * P^T = P * (A + D) * P^T$, applies, which results in the eigenvalues and consequently the characteristic polynomials being equal. □

**Lemma 2.** *Let $A$ and $B$ be two positive definite matrices, and it furthermore, let $\pi$ be a permutation $\pi$ with $\sum_{j=1}^{n}(A_{ji}^k)^2 = \sum_{j=1}^{n}(B_{j\pi_i}^k)^2$ for $k = 1, ..., n$, then they possess identical eigenvalues and consequently identical characteristic polynomials.*

*Proof.* Let $e_1, ..., e_n$ be the unit vectors, let $\mu_1, ..., \mu_m$ be the possibly multiple eigenvalues and $v_{i1}, .., v_{im_i}$ the eigenvectors belonging to $\mu_i$.
$e_i$ can be written as a linear combination of the eigenvectors.
Let now be $e_i = \sum_{l=1}^{m} \sum_{p=1}^{m_l} x_{il_p} v_{lp}$, so applies:
$A^k * e_i = \sum_{l=1}^{m} \left( \mu_l^k * \sum_{p=1}^{m_l} x_{il_p} v_{lp} \right)$ is the $i$th column of $A^k$, called $a_i^{(k)}$. We observe that

$$a_i^{(k)T} a_i^{(k)} = \sum_{l=1}^{m} \left( \mu_l^k * \sum_{p=1}^{m_l} x_{il_p} v_{lp} \right)^T \sum_{l=1}^{m} \left( \mu_l^k * \sum_{p=1}^{m_l} x_{il_p} v_{lp} \right)$$

$$= \left\langle \sum_{l=1}^{m} \mu_l^k * \sum_{p=1}^{m_l} x_{il_p} v_{lp} \;,\; \sum_{l=1}^{m} \mu_l^k * \sum_{p=1}^{m_l} x_{il_p} v_{lp} \right\rangle$$

$$= \sum_{l=1}^{m} \sum_{l'=1}^{m} \mu_l^k \mu_{l'}^k * \sum_{p=1}^{m_l} \sum_{p'=1}^{m_{l'}} x_{il_p} x_{il'_{p'}} \langle v_{l_p}, v_{l'_{p'}} \rangle$$

$$= \sum_{l=1}^{m} \mu_l^{k*2} * \sum_{p=1}^{m_l} x_{il_p}^2 = \sum_{l=1}^{m} \mu_l^{2k} E_{il}$$

$E_{il}^{0.5}$ is thereby the component of the vector $e_i$ that is within the vector space that $\mu_l$ spans.
Once $E_{il}$ are known, we can unambiguously determine the $\mu_l^2$ from the linear system of equations $\sum_{l=1}^{m}(\mu_l^2)^k E_{il} = a_i^{(k)T} a_i^{(k)}$ with $i \in \{1, ..., n\}, k \in \{1, ..., m\}$ according to Vandermonde and, as all eigenvalues are positive, the $\mu_l$ as well. Yet, the $E_{il}$ still have to be identified. The fact that the eigenvectors form an orthonormal basis, yields the condition $\sum_{l=1}^{m} E_{il} = 1$ for $i = 1, ..., m$. This property results in a linear system of equations which is unambiguous given that all values are positive.
As this argumentation equally holds for $B$, both matrices have the same eigenvalues. □

**Gerschgorin's Disc Theorem 1.** *The eigenvalues of a matrix $A = (a_{ij})$ are inside the Gerschgorin discs*

$$G_i = \{z \in C : |z - a_{ii}| \leq \sum_{j=1, j \neq i}^{n} |a_{ij}|\} \quad i = 1, .., n \quad .$$

*Every disc holds an eigenvalue.*

More information about the theorem can be read, for example, in [9]. If $A$ is symmetric, the eigenvalues are real, thus every eigenvalue is in one interval $[a_{ii} - \sum_{j=1, j \neq i}^{n} |a_{ij}|, \; a_{ii} + \sum_{j=1, j \neq i}^{n} |a_{ij}|]$.

**Lemma 3.** *Let $B = D + A$ with $D = diag(d_1, ..., d_n)$ is a diagonal matrix and $A$ is an adjacency matrix of an undirected graph.*

*Let $|d_i - d_j| \geq 4n$ and $d_i > n - 1$ when $d_i \neq d_j$, otherwise $||B^k * e_i||^2 = ||B^k * e_j||^2$ for all $k$. If $d_i = d_j$ then $j$ is in the orbit of $i$, i.e. there is an automorphism $\psi$ with $j = \psi(i)$*

*Proof.* Let $N(w) := \{v | v \text{ is adjacent to vertex } w\}$, then $Be_i = d_i * e_i + \sum_{i' \in N(i)} e_{i'}$. If $d_i = d_j$, the vertices $i$ and $j$ have the same degrees, because $||Be_i||^2 = d_i^2 + \sum_{i' \in N(i)} 1^2 = d_j^2 + \sum_{j' \in N(j)} 1^2 = ||Be_j||^2$. $e_i$ can be written $e_i = \sum_{j=1}^{n} x_{ij} v_j$, where $v_1, ..., v_n$ are eigenvectors, which are an orthonormal basis. Thus

$$1 = \langle e_i, e_i \rangle = \langle \sum_{j=1}^{n} x_{ij} v_j, \sum_{j=1}^{n} x_{ij} v_j \rangle = \sum_{j=1}^{n} \sum_{l=1}^{n} x_{ij} x_{il} \langle v_j, v_l \rangle$$

$$= \sum_{j=1}^{n} x_{ij}^2 \langle v_j, v_j \rangle + \sum_{j \neq l} x_{ij} x_{il} \langle v_j, v_l \rangle = \sum_{j=1}^{n} x_{ij}^2 * 1 + \sum_{j \neq l} x_{ij} * 0 = \sum_{j=1}^{n} x_{ij}^2$$

and
$$e_i^T B e_i = \langle e_i, B e_i \rangle$$
$$= \left\langle \sum_{j=1}^n x_{ij} v_j \ , \ B \sum_{j=1}^n x_{ij} v_j \right\rangle$$
$$= \left\langle \sum_{j=1}^n x_{ij} v_j \ , \ \sum_{j=1}^n B x_{ij} v_j \right\rangle$$
$$= \left\langle \sum_{j=1}^n x_{ij} v_j \ , \ \sum_{j=1}^n \mu_j x_{ij} v_j \right\rangle$$
$$= \sum_{j=1}^n \sum_{l=1}^n \mu_l x_{ij} x_{il} \langle v_j, v_l \rangle$$
$$= \sum_{j=1}^n \mu_j x_{ij}^2 = d_i$$

and it is also
$$(Be_i)^T (Be_i) = \langle Be_i, Be_i \rangle$$
$$= \left\langle B \sum_{j=1}^n x_{ij} v_j \ , \ B \sum_{j=1}^n x_{ij} v_j \right\rangle$$
$$= \left\langle \sum_{j=1}^n B x_{ij} v_j \ , \ \sum_{j=1}^n B x_{ij} v_j \right\rangle$$
$$= \left\langle \sum_{j=1}^n x_{ij} \mu_j v_j \ , \ \sum_{j=1}^n \mu_j x_{ij} v_j \right\rangle$$
$$= \sum_{j=1}^n \sum_{l=1}^n \mu_j \mu_l x_{ij} x_{il} \langle v_j, v_l \rangle$$
$$= \sum_{j=1}^n \mu_j^2 x_{ij}^2 = d_i^2 + \delta_i \ ,$$

where $\delta_i$ is the degree of vertex $i$.
Because of Gerschgorin's disc theorem $B$ is positive definite, thus $e_i$ and $e_j$ are equally distributed to the eigenspaces as shown in Lemma 2, if $d_i = d_j$.
Let $E_{ir} = \sum_{j \in \{j | v_j \in S_r\}} x_{ij}^2$ be the energy of $e_i$ in the eigenspace $S_r = \{v | Bv = \mu_r v\}$, then $\sum_{r=1}^m E_{ir} = 1$. We can look at $E_{i1}, ..., E_{im}$ as if it is a probability distribution.
In this case $E[\mu_r] = \sum_{r=1}^m \mu_r E_{ir} =< e_i, Be_i >= d_i$ and $Var[\mu_r] = E[\mu_r^2] -$

$E[\mu_r]^2 = \sum_{r=1}^{m} \mu_r^2 E_{ir} - [\sum_{r=1}^{m} \mu_r E_{ir}]^2 = <Be_i, Be_i> - <e_i, Be_i> = d_i^2 + \delta_i - d_i^2 = \delta_i$. By Chebyshev's inequality there will be $Pr(|\mu_r - d_i| \geq 2n) \leq \frac{\delta_i}{4n^2} < \frac{n}{4n^2} = \frac{1}{4n}$. If an eigenvalue $\mu_r$ is outside the Gerschgorin disc, then $E_{ir} < \frac{1}{4n}$.

Let $e_i$ and $e_j$ two vectors with $\forall r E_{ir} = E_{jr}$, so the energy of $Be_i$ and $Be_j$ is also equaly distributed to the eigenspaces and so $\forall r \sum_{i' \in N(i)} E_{i'r} = \sum_{j' \in N(j)} E_{j'r}$. For any $d$ let $k_i$ the number of vertices $i' \in N(i)$ with $d_{i'} = d$ and $k_j$ the number of vertices $j' \in N(j)$ with $d_{j'} = d$. $k_i = k_j$, because :

Assumption: $k_j < k_i$.

Let $G(d)$ be the Gerschgorin disk $[d - 2n, d + 2n]$.

Let $E_{G(d)i} = \sum_{r \in \{r | \mu_r \in G(D)\}} E_{ri}$ the energy in the eigenspaces with eigenvalues inside the Gerschgorin disc, then

$$\sum_{i' \in N(i)} E_{G(d)i'} \geq k_i(1 - \frac{1}{4n}) > k_i - \frac{1}{4}$$

and

$$\sum_{j' \in N(j)} E_{G(d)j'} \leq k_j + (N(j) - k_j)\frac{1}{4n} < k_j + \frac{1}{4} \quad .$$

From that follows $k_i = k_j$.

So there is a bijection between adjacent vertices of $i$ and $j$ with $l$ and $\psi(l)$ with $d_l = d_{\psi(l)}$. This way we find a bijection on every vertex of the graph, which is described by a permutation matrix $P$. If $i$ and $j$ are vertices with $e_j = Pe_i$, then $B_{ii} = d_i = d_j = B_{jj}$ and so $E_{ir} = E_{jr}$ for every $r \in \{1, ..., m\}$. Let $e_i = \sum_{r=1}^{m} u_{ir}$ and $e_j = \sum_{r=1}^{m} u_{jr}$, where $u_{ir}$ and $u_{jr}$ are eigenvectors in the eigenspace $S_r$, then $||u_{ir}||^2 = E_{ir} = E_{jr} = ||u_{ir}||^2$.

$\forall r \ u_{jr} = Pu_{ir}$, because $A^k P \sum_r \mu_r^k u_{jr} = A^k Pe_i = A^k e_j = \sum_r \mu_r^k u_{jr}$ for every $k$ and $\mu_r > 0$ are distinct eigenvalues, thus a linear system of equations with exactly one solution is build.

This causes $PBe_i = P * \sum_{r=1}^{m} \mu_r u_{ir} = \sum_{r=1}^{m} \mu_r P * u_{ir} = \sum_{r=1}^{m} \mu_r u_{jr} = B * e_j = B * Pe_j$, thus $PB = BP$. For the adjacency matrix $A = B - diag(d_1, ..., d_n)$ we get $PA = AP$ and so $P$ is an automorphism, as proven in [5]. □

## 4 Algorithms

### 4.1 Algorithm 1

**input :** adjacency matrices of graphs $A$ und $A'$
**output:** graphs isomorphic ?

$C1_1 := \{1, .., n\}$
$C2_1 := \{1, .., n\}$
$A1 := A$
$A2 := A'$

```
repeat
    c := number of C1_x
    for i:=1 to n
        A1[i,i] := 4n * x, where i ∈ C1_x
        A2[i,i] := 4n * x, where i ∈ C2_x
    for k:=1 to n
        for i:=1 to n
            a1_i^(k) := ith column vector of A1^k
            a2_i^(k) := ith column vector of A2^k
            Â1[i][k] := a1_i^(k)T a1_i^(k)
            Â2[i][k] := a2_i^(k)T a2_i^(k)
    sort lists Â1 and Â2 lexicographically
    compare Â1 and Â2
    if sorted lists are not equal
        then return false
    put vertices into equivalence classes C1_x, such that i ≡ j iff Â1[i] = Â1[j]
    put vertices into equivalence classes C2_x, such that i ≡ j iff Â2[i] = Â2[j]
        and Â1[l] = Â2[i] iff l ∈ C1_x and i ∈ C2_x
until c = number of C_x
return true
```

If the graphs are isomorph, the algorithm will return true, because not only the adjacent matrices have the same spectra, also the matrices A1 ad A2 have the same spectra in every iteration step because of Lemma 1.

If Algorithm 1 returns true, we can look at the graph as two not connected parts of a big graph. Because of Lemma 3 we find for every vertex in the first graph a vertex in the second graph, with is in the same orbit, belonging to the big graph. That way, the algorithm construct an isomorphism.

### 4.2 Algorithm 2

Algorithm to construct an isomorphism

**input :**  adjacency matrices of two isomorphic graphs $A$ und $A'$
**output:**  graph isomorphism, if exists

```
make A1,A2,C1,C2 and c as described in Algorithm 1
while c < n
    get x with |C1_x| = |C2_x| > 1
    get an i ∈ C1_x and j ∈ C2_x
    A1[i,i] := A2[j,j] := 4 * (c+1) * n
    for k:=1 to n
        for i:=1 to n
```

$$a1_i^{(k)} := i\text{th column vector of } A1^k$$
$$a2_i^{(k)} := i\text{th column vector of } A2^k$$
$$\hat{A}1[i][k] := {a1_i^{(k)}}^T a1_i^{(k)}$$
$$\hat{A}2[i][k] := {a2_i^{(k)}}^T a2_i^{(k)}$$

    put vertices into equivalence classes $C1_x$, such that $i \equiv j$ iff $\hat{A}1[i] = \hat{A}1[j]$
    put vertices into equivalence classes $C2_x$, such that $i \equiv j$ iff $\hat{A}2[i] = \hat{A}2[j]$
      and $\hat{A}1[l] = \hat{A}2[i]$ iff $l \in C1_x$ and $i \in C2_x$
  $c :=$ number of $C1_x$
 sort lists $\hat{A}1$ and $\hat{A}2$ lexicographically
 return permutation induced by the sorted lists

When $c = n$, we have only single eigenvalues for $A1$ and $A2$. In this case, there is only one solution for $A1 = P * A2 * P^T$ with $P$ unitary, the permutation the algorithm has produced. Now $A = P * A' * P^T$ because of Lemma 1 and so the permutation is an isomorphism of the graphs.

## 5 Complexity

The main effort of Algorithm 1 consists in $n$ matrix multiplications in every iteration of the outer loop. The outer loop is a fixed point iteration, because it terminates, when no more equivalence classes can be constructed. In case of constructing new classes, the number of classes will grow. Algorithm 1 must not make more then $n$ iterations in the outer loop.
In the outer loop of Algorithm 2 vertex $i$ of the first graph and vertex $j$ of the second graph will put into a single class separated to the other vertices. Algorithm 2 also have not more then $n$ iterations in outer loop.
Both algorithms consists in $n^2$ matrix multiplications in $n$ iterations of the outer loop. Thus we have to perform $O(n^8)$ integer additions or multiplications.
The maximum eigenvalue of $A1$ is according to Gerschgorin's circle theorem $\leq 4\frac{n(n+1)}{2} = 2n^2 + 2n$. Consequently the maximum eigenvalue of $A1^n \leq (2n^2 + 2n)^n$. Since all matrix entries are positive, none of them can be bigger than the maximum eigenvalue and therefore $\leq (2n^2 + 2n)^n$. The maximum value occurs in calculating the sum of the quadratic entries and is $\leq n * (2n^2 + 2n)^{2n}$. The coding length for an entry is therefore $(6n) * (\log(n) + C) = O(n * \log(n))$.
Algorithm 2 needs also at most $O(n^2)$ matrix multiplications with maximum coding length of an entry in $O(n * \log(n))$.

## 6 Literature

1. McKay, Brendan D. (1981), Practical graph isomorphism, Congressus Numerantium 30: 10th. Manitoba Conference on Numerical Mathematics and Computing (Winnipeg, 1980)